\documentclass[showpacs,preprintnumbers,amsmath,amssymb,12pt]{article}
\usepackage{graphicx,amsfonts,amssymb,amsthm,amsmath,graphics,epsfig,pstricks,fancyhdr,fancybox}
\usepackage{dcolumn}
\usepackage{bm}

\def\RE{{\bm R}}
\newcommand{\refeq}[1]{~(\ref{#1})}

\newcommand{\pdf}{\textit{pdf}}
\newcommand{\cdf}{\textit{cdf}}

\newcommand{\PR}[1]{\bm{P}\left\{{#1}\right\}}

\newcommand{\norm}{\mathfrak{N}}

\newcommand{\poiss}{\mathfrak{P}}

\newcommand{\cauchy}{\mathfrak{C}}
\newcommand{\esp}{\mathfrak{E}}
\newcommand{\gam}{\mathfrak{G}}
\newcommand{\lapl}{\mathfrak{L}}
\newcommand{\unif}{\mathfrak{U}}

\newcommand{\stud}{\mathfrak{T}}
\newcommand{\bin}{\mathfrak{B}}

\newcommand{\ton}{\overset{n}{\longrightarrow}}

\begin{document}
\thispagestyle{empty}

\title{\Huge \textbf{Entropy and its discontents:}\\
\Huge \textbf{A note on definitions}}
\author{\textsc{Nicola Cufaro Petroni}\\
Dipartimento di Matematica and \textsl{TIRES}, Universit\`a di Bari\\
\textsl{INFN} Sezione di Bari\\
via E. Orabona 4, 70125 Bari, Italy\\
\textit{email: cufaro@ba.infn.it}}

\date{}

\maketitle

\begin{abstract}
\noindent The routine definitions of Shannon entropy for both
discrete and continuous probability laws show inconsistencies that
make them not reciprocally coherent. We propose a few possible
modifications of these quantities so that 1) they no longer show
incongruities, 2) they go one into the other in a suitable limit as
the result of a renormalization. The properties of the new
quantities would slightly differ from that of the usual entropies in
a few other respects
\end{abstract}




\section{Introduction}\label{intro}

As it is usually defined, the \emph{Shannon entropy} of a discrete
law $p_k=\PR{x_k}$ associated to the values $x_k$ of some random
variable is
\begin{equation}\label{entropy}
    H=-\sum_kp_k\ln p_k
\end{equation}
and apparently is a non negative, dimensionless quantity. As a
matter of fact however it does not depend on all the details of the
distribution: for instance only the $p_k$ are relevant, while the
$x_k$ play no role at all. This means that if we modify our
distribution just by moving the $x_k$, the entropy is left the same:
this entails among others that $H$ does not always change along with
the \emph{variance} (or other typical parameters) of the
distribution which instead is contingent on the $x_k$ values. In
particular $H$ is invariant under every linear transformation
$ax_k+b$ (centering and rescaling) of the random quantities: in this
sense every \textit{type of laws}~\cite{loeve} is isoentropic.
Surprisingly enough, despite the unsophistication of
definition\refeq{entropy}, and beyond a few elementary examples,
explicit formulas displaying the dependence of the entropy $H$ from
the parameters of the most common discrete distributions are not
known. If for instance we take the entropy $H$ of the binomial
distributions $\bin_{n,p}$ with
\begin{equation}\label{binomial}
    x_k=k=0,1,\ldots,n\qquad\quad p_k=\binom{n}{p} p^k(1-p)^{n-k}
\end{equation}
although it would always be possible to calculate the entropy $H$
for every particular example, no general formula giving its explicit
dependence from $n$ and $p$ is available, and only its asymptotic
behavior for large $n$ is known in the
literature~\cite{jacquet,cichon}:
\begin{equation}\label{asymptbinomial}
    H\left[\bin_{n,p}\right]=\frac{1}{2}\,\ln\left[2\pi
    enp(1-p)\right]+\frac{4p(1-p)-1}{12np(1-p)}+O\left(\frac{1}{n^2}\right)
\end{equation}
Remark moreover that, while this formula explicitly contains
$np(1-p)$ namely the variance of $\bin_{n,p}\,$, it is easy to
recognize that -- as long as we leave untouched the $n$
probabilities $p_k$ -- the entropy $H\left[\bin_{n,p}\right]$
remains the same when we change the variance by moving the points
$x_k$ away from their usual locations $x_k=k$. In particular this is
true for the standardized (centered, \emph{unit variance}) binomial
$\bin_{n,p}^*$ with
\begin{equation}\label{stbinomial}
    x_k=\frac{k-np}{\sqrt{np(1-p)}}\qquad\quad k=0,1,\ldots,n
\end{equation}
and same $p_k$ of\refeq{binomial} which entails
$H\left[\bin_{n,p}\right]=H\left[\bin_{n,p}^*\right]$. All that
hints to the fact that what seems to be relevant to the entropy is
not the variance itself, but some other feature, possibly related to
the shape of the distribution. In a similar vein, for the Poisson
distributions $\poiss_\lambda$ with
\begin{equation}\label{poisson}
    x_k=k=0,1,\ldots\qquad\quad p_k=e^{-\lambda}\frac{\lambda^k}{k!}
\end{equation}
the entropy is
\begin{equation}\label{poissentropy}
     H\left[\poiss_\lambda\right]=\lambda(1-\ln\lambda)+e^{-\lambda}\sum_{k=0}^\infty\frac{\lambda^k\ln k!}{k!}
\end{equation}
with an asymptotic expression for large $\lambda$
\begin{equation}\label{asymptpoisson}
     H\left[\poiss_\lambda\right]=\frac{1}{2}\,\ln\left(2\pi
     e\lambda\right)-\frac{1}{12\lambda}-\frac{1}{24\lambda^2}-\frac{19}{360\lambda^3}+O\left(\lambda^{-4}\right)
\end{equation}
which explicitly contains the parameter $\lambda$ (also playing the
role of the variance), but which is also completely independent from
the values of the $x_k$'s. As a consequence a standardized Poisson
distribution $\poiss^*_\lambda$, with
\begin{equation}\label{stpoisson}
    x_k=\frac{k-\lambda}{\sqrt{\lambda}}
\end{equation}
and same probabilities $p_k$, has the same entropy of
$\poiss_\lambda$, namely
$H\left[\poiss_\lambda\right]=H\left[\poiss^*_\lambda\right]$.

When on the other hand we consider continuous laws\footnote{For
short we will call \emph{continuous} the laws possessing a \pdf\
$f(x)$, without insisting on the difference between
\emph{continuous} and \emph{absolutely continuous} distributions
which is not relevant here.} with a \pdf\ $f(x)$, the
definition\refeq{entropy} no longer apply, and we are led to
introduce another quantity commonly known as \emph{differential
entropy} (we acknowledge that this name could be misleading for an
integral, but we will retain it in the following to abide a long
established habit):
\begin{equation}\label{diffentropy}
    h=-\int_\RE f(x)\ln f(x)\,dx
\end{equation}
which in several respects differs from the entropy\refeq{entropy} of
the discrete distributions. First of all explicit formulas of the
entropy\refeq{diffentropy} are known for most of the usual laws: for
example (see also Appendix~\ref{examples}) the distributions
$\unif(a)$ uniform on $[0,a]$ with $a>0$ have entropy
\begin{equation}\label{uniformentropy}
    h\left[\unif\,\right]=\ln a
\end{equation}
while for the centered, Gaussian laws $\norm(a)$ with variance $a^2$
we have
\begin{equation}\label{normalentropy}
    h\left[\norm\,\right]=\ln\left(a\sqrt{2\pi e}\right)
\end{equation}
An exhaustive list of similar formulas for other families of laws is
widely available in the literature, but even from these two examples
only it is apparent that:
\begin{enumerate}
    \item at variance with the discrete case, the differential entropies explicitly depend
    on a scaling parameter $a$, showing now a dependence either on the variance, or on some other dispersion index such as the \emph{inter quantile
    ranges} (\emph{IQnR}): this means in particular that the types of continuous laws are no
    longer isoentropic;
    \item the differential entropies can take negative values when the parameters of the laws are chosen in such a way that the logarithm arguments fall
    below 1;
    \item the logarithm arguments are not in general dimensionless
    quantities, in an apparent violation of the homogeneity rule that the scalar arguments of transcendental
    functions (as logarithms are) must be dimensionless quantities;
    this entails in particular that the entropy depends on the
    units of measurement
\end{enumerate}
These three remarks make hence abundantly clear that something is
inscribed in the definition\refeq{diffentropy} which is not present
in the definition\refeq{entropy}, and vice versa.

Finally the two definitions seem not to be reciprocally consistent
in the sense that, when for instance a continuous law is weakly
approximated by a sequence of discrete laws, we would like to see
the entropies of the discrete distributions converging toward the
entropy of the continuous one. That this is not the case is apparent
from a few counterexamples. It is well known for instance that, for
every $0<p<1$, the sequence of the standardized binomial laws
$\bin_{n,p}^*$ weakly converges to the Gaussian $\norm(1)$ when
$n\to\infty$; however, since the binomial probabilities $p_k$ are
unaffected by a standardization, the entropies
$H\left[\bin_{n,p}^*\right]$ still obey to
formula\refeq{asymptbinomial}, and hence their sequence diverges as
$\ln \sqrt{n}$ instead of being convergent to the differential
entropy of $\norm(1)$ which from\refeq{normalentropy} is
$\ln\sqrt{2\pi e}$. In the same vein the \cdf\ $F(x)$ of a uniform
law $\unif(a)$ can be approximated by the sequence $F_n(x)$ of the
discrete uniform laws $\unif_n(a)$ concentrated with equal
probabilities $p_1=\ldots=p_n=\frac{1}{n}$ on the $n$ equidistant
points $x_1,\ldots,x_n$ where $x_k=k\Delta$ for $k=1,2,\ldots,n$,
and $x_k-x_{k-1}=\Delta=\frac{a}{n}$ with $x_0=0$. However it is
easy to see that
\begin{equation}\label{discrunifentropy}
    H[\unif_n]=-\sum_{k=1}^n\frac{1}{n}\ln\frac{1}{n}=\ln n
\end{equation}
so that their sequence again diverge as $\ln n$, while the
differential entropy $h[\unif\,]$ of the uniform law has the finite
value\refeq{uniformentropy}.

As a consequence of these remarks, in the following sections we will
propose a few elementary ways to change the two
definitions\refeq{entropy} and\refeq{diffentropy} in order to
possibly rid them of the said inconsistencies, and to make them
reciprocally coherent without losing too much of the essential
properties of the usual quantities. These new definitions, moreover,
operate an effective renormalization of the said divergences so that
now, when a continuous law is weakly approximated by a sequence of
discrete laws, also the entropies of the discrete distributions
converge toward the entropy of the continuous one. A few additional
points with examples and explicit calculations are finally collected
in the appendices. It must be clearly stated at this point, however,
that we do not claim here that the Shannon entropy is somehow
ill-defined in itself: we rather point out a few reciprocal
inconsistencies of the different manifestations of this time-honored
concept, and we try to attune them in such a way that every
probability distribution (either discrete, or continuous) would now
be treated on the same foot


\section{Entropy for continuous laws}\label{contlaws}

Let us begin with some remarks about the differential entropy for
continuous laws with a \pdf\ $f(x)$: the simplest ways to achieve
the essential of our aims would be to adopt some new definition of
the type
\begin{equation}\label{newdiffentropy2}
    -\int_\RE f(x)\ln\left[\kappa
    f(x)\right]\,dx=h-\ln\kappa
\end{equation}
where $\kappa$ is any parameter of the law $f(x)$ with the same
dimensions of $x$, and with a finite and strictly positive value for
every non degenerate law. To this end the first idea which comes to
the fore consists in taking the \emph{standard deviation} $\sigma$
to play the role of $\kappa$ in\refeq{newdiffentropy2}, but it is
also apparent that this choice would restrict our definition only to
the continuous laws \emph{with finite second momentum} leaving out
many important cases. A strong alternative candidate for the role of
$\kappa$ could instead be some \emph{interquantile range}
(\emph{IQnR}) which can represent a measure of the dispersion even
when the variance does not exist. In the following we will analyze a
few possible choices for the parameter $\kappa$ along with their
principal consequences

\subsection{Interquantile range}\label{interquant}

The calculation of the \emph{IQnR} goes through the use of the
\emph{quantile function} $Q(p)$, namely the inverse \emph{cumulative
distribution function} (\cdf). In order to take into account
possible jumps and flat spots of a given \cdf\ $F(x)$, the quantile
function is usually defined as
\begin{equation}\label{qf1}
    Q(p)=\inf\{x\in\RE :p\leq F(x)\}\qquad\quad0\leq
p\leq1
\end{equation}
In the case of continuous laws (no jumps), however, this can be
reduced to
\begin{equation}\label{qf2}
    Q(p)=\inf\{x\in\RE :p= F(x)\}
\end{equation}
and when $F(x)$ is also strictly increasing (no flat spots) we
finally have
\begin{equation}\label{qf3}
    Q(p)=F^{-1}(p)
\end{equation}
It is apparent then that $Q(p)$ jumps wherever $F(x)$ has flat
spots, while it has flat spots wherever $F(x)$ jumps. The
\emph{IQnR} function is then defined as
\begin{equation}\label{iqnr}
    \varrho(p)=Q(1-p)-Q(p)\qquad\quad0< p<\frac{1}{2}
\end{equation}
and the classical \emph{interquartile range} (\emph{IQrR}) is just
the particular value
\begin{equation}\label{iqr}
    \varrho\left(\frac{1}{4}\right)=Q\left(\frac{3}{4}\right)-Q\left(\frac{1}{4}\right)
\end{equation}
The \emph{IQnR} $\varrho(p)$ is a non increasing function of $p$,
and for continuous laws (since $Q(p)$ has no flat spots) it is
always well defined and never vanishes so that one of its values can
be safely used to play a role in the definition of $\kappa$
in\refeq{newdiffentropy2}. Of course, when a law has also a finite
second momentum, the \emph{IQnR} $\varrho$ and the standard
deviation $\sigma$ are both well defined and the ratio
$\gamma=\varrho/\sigma$ often has the same for entire families of
laws. We now propose to adopt a new form for the entropy of
continuous laws which -- by making use, instead of the variance, of
some particular value of \emph{IQnR} that we will denote
$\widetilde{\varrho}$ -- will encompass even the case of the laws
without a finite second momentum:
\begin{equation}\label{newestdiffentropy}
    \widetilde{h}=-\int_\RE f(x)\ln\left[\widetilde{\varrho}
    \,f(x)\right]\,dx=h-\ln\widetilde{\varrho}
\end{equation}
In particular for the continuous laws we can simply take
$\widetilde{\varrho}=\varrho\left(\frac{1}{4}\right)$, the
\emph{IQrR}

Despite the minimality of this change of definition, however, the
new entropy $\widetilde{h}$ has properties slightly different from
$h$. It is shown by the examples of the Appendix~\ref{examples} that
-- at variance with the usual differential entropy $h$ -- this new
entropy $\widetilde{h}$ has neither a minimum, nor a maximum value
because, according to the particular continuous law considered, it
takes every possible real value, both positive and negative. In this
respect we must instead recall the well known property of the
Gaussian laws $\norm(a)$ which qualify as the laws with the maximum
differential entropy $h$ among all the other continuous laws with
the same variance $\sigma^2$. It is apparent then that within our
new definition\refeq{newestdiffentropy} this special position of the
Gaussian laws will simply be lost.

The adoption of\refeq{newestdiffentropy} however brings several
benefits that will also be made apparent in the examples of the
Appendix~\ref{examples}: first of all the argument of the logarithm
is now by definition a \emph{dimensionless} quantity so that the
value of $\widetilde{h}$ becomes invariant under change of
measurement units. Second, the new entropy $\widetilde{h}$ will no
longer depend on the value of some scaling parameters linked to the
variance: its values are determined by the \emph{form of the
distribution}, rather than by its actual numerical dispersion, and
will be the same for entire families of laws. When in fact the
variables are subject to some linear transformation $y=ax+b$ (with
$a>0$, as in the changes of unit of measurement) the differential
entropy $h$ changes with the new \emph{pdf} according to
\begin{equation*}
   -\int_\RE \frac{1}{a}f\left(\frac{y-b}{a}\right)\ln\left[
   \frac{1}{a}f\left(\frac{y-b}{a}\right)\right]\,dy=-\int_\RE
   f(x)\ln f(x)\,dx+\ln a
\end{equation*}
namely it is explicitly dependent from the scaling parameter $a$,
while it is independent from the centering parameter $b$. It is
apparent moreover that, according to these remarks, also the
quantile function of the transformed \emph{cdf}
\begin{equation*}
    F\left(\frac{x-b}{a}\right)
\end{equation*}
is changed into $aF^{-1}(p)+b=aQ(p)+b$ so that any \emph{IQnR} is
modified according to $a\varrho(p)$, namely it will be sensitive
again only to the scaling parameter $a$, but not to the centering
one. As a consequence the modifications of both $h$ and
$\widetilde{\varrho}$ under a linear transformation of the variables
are apparently such that they cancel out reciprocally so that
$\widetilde{h}$, as defined in\refeq{newestdiffentropy}, is always
left unchanged: this means in particular that \emph{the types of
laws are isoentropic}

\subsection{Variance and scaling parameters}\label{alternative}

By restricting ourselves to the continuous laws \emph{with finite
second momentum} and standard deviation $\sigma$, an alternative
redefinition of the differential entropy could be considered as
\begin{equation}\label{newdiffentropy3}
    -\int_\RE f(x)\ln\left[\sigma
    f(x)\right]\,dx=h-\ln\sigma
\end{equation}
This form of differential entropy would bring the same benefits of
$\widetilde{h}$: the argument of the logarithm is dimensionless, and
it will no longer depend on the value of scaling parameters. Since
however the dimensional parameter is the standard deviation, it is
possible to show that the Gaussian laws would keep now their usual
role of \emph{maximum entropy} laws, and this suggests to propose a
further possible change of definition as (please remark the
\emph{change of sign})
\begin{equation}\label{newdiffentropy}
    \widehat{h}=\int_\RE f(x)\ln\left[\sigma\sqrt{2\pi
    e}\,f(x)\right]\,dx=\ln\left(\sigma\sqrt{2\pi e}\right)-h
\end{equation}
As shown in the Appendix~\ref{examples}, all the Gaussian laws
$\norm(a)$ will now have $\widehat{h}[\norm]=0$ and -- because of
the change of sign -- this value will now represent the
\emph{minimum} for all the other laws, irrespective of their
variance: as a consequence the entropy $\widehat{h}$ of all the
continuous distributions with finite variance will now be \emph{non
negative}, as for the entropy $H$ of the discrete laws.

For laws lacking a finite second momentum (as the Cauchy laws) we
would have no $\widehat{h}$ entropy because these distributions have
no variance to speak about: this is an apparent shortcoming
presented by the definition\refeq{newdiffentropy} of $\widehat{h}$,
and to go around this weakness we introduced our
definition\refeq{newestdiffentropy} of $\widetilde{h}$ by exploiting
the properties of the \emph{IQnR} $\varrho(p)$ which are always well
defined for every possible distribution. It would be interesting to
remark, however, not only that these are not the only two possible
choices, but also that even seeming harmless modifications can imply
slightly different properties. For instance, by going back to the
remarks at the end of the Section~\ref{interquant}, it is well known
that by linear transformation of the variables (with $a>0$ to
simplify) every continuous law $f(x)$ spans a \emph{type} of
continuous laws
\begin{equation*}
    \frac{1}{a}f\left(\frac{x-b}{a}\right)
\end{equation*}
As already pointed out, the centering parameter $b$ has no influence
on the value of the entropy, while the \emph{scaling parameter} $a$
would change the differential entropy $h$ of
definition\refeq{diffentropy} by an additional $\ln a$. As a
consequence by simply adopting as a new definition
\begin{equation}\label{typentropy}
    \overline{h}=-\int_\RE f(x)\ln\left[af(x)\right]\,dx=h-\ln a
\end{equation}
where $a$ is the parameter locating the law within its type, we
would get an entropy invariant for rescaling. It is apparent that
the definition\refeq{typentropy} considers $a$ just as a parameter,
and not as a measure of dispersion, and it is interesting to notice
that it also entails a few consequences shown in the examples of the
Appendix~\ref{examples}. In particular we now have that the entropy
$\overline{h}$ takes again all the (positive and negative) real
values and hence that there is no such a thing as a maximum entropy
distribution as in the case of the $\widetilde{h}$ entropy

\section{Entropy for discrete laws}\label{discrlaws}

We could now naively extend to the discrete laws our previous
re-definitions simply by taking $H-\ln\kappa$, with $H$ given
by\refeq{entropy}, and with a suitable choice of $\kappa$, but in so
doing we would miss a chance to reconcile the two forms (discrete
and continuous) of our entropy in some limit behavior. We find then
more convenient to introduce some further changes that for the sake
of generality we will discuss in the settings of the
Section~\ref{interquant} where $\kappa$ is an \emph{IQnR}

\subsection{Renormalization}

In order to extend the definition\refeq{newestdiffentropy} to the
discrete distributions we must first remark that, at variance with
the continuous case, now the \emph{IQrR}
$\varrho\left(\frac{1}{4}\right)$ can vanish and hence can not be
immediately adopted as $\kappa$ in our definitions. For the discrete
laws in fact $F(x)$ makes jumps and hence $Q(p)$ has flat spots so
that $\varrho(p)$ can be zero for some values of $p$, and in
particular this can happen also for $p=\frac{1}{4}$. If however our
distributions are purely discrete (a few remarks about the more
general case of mixtures can be found in the
Appendix~\ref{mixtures}) $\varrho(p)$ is a non increasing function
of $p$ which change values only by jumping, and which is constant
between subsequent jumps. As a consequence, with the only exception
of the degenerate laws (which have a constant $Q(p)$, and hence a
$\varrho$ vanishing for every $p$), $\varrho(p)$ certainly takes non
zero values for some $0< p\leq\frac{1}{4}$, even when
$\varrho\left(\frac{1}{4}\right)=0$. We can then use in our
definitions as dimensional constant $\widetilde{\varrho}$ the
smallest, non zero \emph{IQnR} larger or equal to the \emph{IQrR}
$\varrho\left(\frac{1}{4}\right)$: more precisely, if $\mathcal{P}$
is the set of all the values of $\varrho(p)$ for $0<
p\leq\frac{1}{4}$, and $\mathcal{P}_0=\mathcal{P}\backslash \{0\}$,
we will take $\widetilde{\varrho}=\min\mathcal{P}_0>0$. Remark that
in particular we again have
$\widetilde{\varrho}=\varrho\left(\frac{1}{4}\right)$ whenever the
\emph{IQrR} does not vanish.

We start by remarking that if $F(x)$ is the cumulative distribution
function of a discrete distribution concentrated on $x_k$ with
probabilities $p_k$ for $k=1,2\ldots\,$, by taking
\begin{equation*}
    \Delta x_k=x_k-x_{k-1}\qquad\quad\Delta
    F_k=F(x_k)-F(x_{k-1})=p_k\qquad\quad k=1,2,\ldots
\end{equation*}
(with $x_0<x_1$ and hence $F(x_0)=0$: for instance
$x_0=x_1-\inf_{k\geq2}\Delta x_k$ so that $\Delta
x_1=\inf_{k\geq2}\Delta x_k$) we see first that the
definition\refeq{entropy} can be immediately recast in the form
\begin{equation*}
    H=-\sum_{k\geq1}\Delta F_k\ln\Delta F_k
\end{equation*}
Since on the other hand many typical discrete distributions
(binomial, Poisson $\ldots$) describe counting experiments, in many
instances we have $\Delta x_k=1$ and in these cases (since $\Delta
x_k$ is also dimensionless) we could also write
\begin{equation*}
    H=-\sum_{k\geq1}\Delta F_k\ln\frac{\Delta F_k}{\Delta x_k}
\end{equation*}
By comparing this expression with the definition of differential
entropy\refeq{diffentropy}, and by recalling that for a continuous
distribution we have $f(x)=F\,'(x)$, we are led to propose as new
definition of the entropy of a discrete law the quantity
\begin{equation}\label{newentropy}
    \widetilde{H}=-\sum_{k\geq1}\frac{\Delta F_k}{\Delta x_k}\ln\left(\widetilde{\varrho}\,\frac{\Delta F_k}{\Delta x_k}\right)\Delta x_k=H+\sum_{k\geq1}p_k\ln\Delta
    x_k-\ln\widetilde{\varrho}
\end{equation}
In general, even for the discrete distributions, $\Delta x_k$ is not
dimensionless, but apparently this is compensated by means of
$\widetilde{\varrho}$. This definition\refeq{newentropy} has
properties which are similar to that of the new differential entropy
defined in the previous section, but the main benefit of this new
formulation is that now -- as it will be discussed in the subsequent
section -- the differential $\widetilde{h}$ entropy of a continuous
law\refeq{newestdiffentropy} can be recovered as a limit of the
entropies $\widetilde{H}$ of a sequence of approximating, discrete
laws. In fact the new definition\refeq{newentropy} effectively
\emph{renormalizes} the traditional entropy $H$ in such a way that
the asymptotic divergences pointed out in the Section~\ref{intro}
are exactly compensated by means of our dimensional parameters.
These conclusions hold also for a suitable extension of the
alternative definitions\refeq{newdiffentropy} and\refeq{typentropy}
respectively of $\widehat{h}$ and $\overline{h}$

\subsection{Convergence}

We will discuss in this section a few particular examples showing
that the two quantities $\widetilde{h}$ of\refeq{newestdiffentropy}
and $\widetilde{H}$ in\refeq{newentropy} are no longer disconnected
concepts, as happens for the usual definitions\refeq{entropy}
and\refeq{diffentropy}. Let us consider first the case of the
binomial laws $\bin_{n,p}$ and of their standardized versions
$\bin_{n,p}^*$ already introduced in the Section~\ref{intro}: we
know that $H[\bin_{n,p}]=H[\bin_{n,p}^*]$ and that both these
entropies diverge as $\ln\sqrt{n}$ when $n\to\infty$. On the other
hand, since $\Delta x_k=1$ for $\bin_{n,p}$, from\refeq{newentropy}
we get
\begin{equation*}
    \widetilde{H}[\bin_{n,p}]=H[\bin_{n,p}]-\ln\widetilde{\varrho}
\end{equation*}
For binomial laws with fixed $p$, and $n$ large enough, the
\emph{IQrR} does not vanish so that
$\widetilde{\varrho}=\varrho\left(\frac{1}{4}\right)$ and, by
introducing the ratio
\begin{equation}\label{gamma}
    \gamma=\frac{\varrho\left(\frac{1}{4}\right)}{\sigma}
\end{equation}
we have
\begin{equation*}
    \widetilde{\varrho}=\gamma\sigma=\gamma[\bin_{n,p}]\sqrt{np(1-p)}
\end{equation*}
and hence
\begin{equation*}
    \widetilde{H}[\bin_{n,p}]=H[\bin_{n,p}]-\ln\widetilde{\varrho}=H[\bin_{n,p}]-\ln\left(\gamma[\bin_{n,p}]\sqrt{np(1-p)}\right)
\end{equation*}
From\refeq{asymptbinomial} for large $n$, since from the binomial
limit theorem we have $\gamma[\bin_{n,p}]\ton\gamma[\norm]$, while
from\refeq{gauss} and\refeq{gauss3} in Appendix~\ref{examples} it is
\begin{equation*}
    \gamma[\norm]=\Phi^{-1}\left(\frac{3}{4}\right)-\Phi^{-1}\left(\frac{1}{4}\right)
\end{equation*}
by taking into account\refeq{gausstilde} of Appendix~\ref{examples},
we finally get
\begin{eqnarray*}
    \widetilde{H}[\bin_{n,p}]&=&\frac{1}{2}\,\ln\left[2\pi
    enp(1-p)\right]+\frac{4p(1-p)-1}{12np(1-p)}+O\left(\frac{1}{n^2}\right)-\ln\left[\gamma[\bin_{n,p}]\sqrt{np(1-p)}\right]\\
    &=&-\ln\left(\frac{\gamma[\bin_{n,p}]}{\sqrt{2\pi
    e}}\right)+\frac{4p(1-p)-1}{12np(1-p)}+O\left(\frac{1}{n^2}\right)\ton-\ln\left(\frac{\gamma[\norm]}{\sqrt{2\pi
    e}}\right)=\widetilde{h}[\norm]
\end{eqnarray*}
which is a first example of the convergence of entropies to
differential entropies in the framework of the new definitions. The
same result is achieved for $\widetilde{H}[\bin_{n,p}^*]$ because
now $\sigma=1$, so that $\widetilde{\varrho}=\gamma[\bin_{n,p}]$
while from\refeq{stbinomial} we have for every $k$
\begin{equation*}
    \Delta x_k=\frac{1}{\sqrt{np(1-p)}}
\end{equation*}
and hence from\refeq{newentropy} we get
\begin{equation*}
    \widetilde{H}[\bin_{n,p}^*]=H[\bin_{n,p}^*]-\ln\left(\gamma[\bin_{n,p}^*]\sqrt{np(1-p)}\right)
\end{equation*}
On the other hand we know that $H[\bin_{n,p}^*]=H[\bin_{n,p}]$, so
that from $\gamma[\bin_{n,p}^*]\ton\gamma[\norm]$ we get again
\begin{equation*}
    \widetilde{H}[\bin_{n,p}^*]\ton\widetilde{h}[\norm]
\end{equation*}
Similar results hold for the Poisson laws $\poiss_\lambda$: we
already remarked in the Section~\ref{intro} that, while
$\poiss^*_\lambda\to\norm(1)$ for $\lambda\to\infty$, the entropies
$H[\poiss_\lambda]=H[\poiss^*_\lambda]$ diverge as
$\ln\sqrt{\lambda}$. It could be shown instead that both
$\widetilde{H}[\poiss_\lambda]$ and
$\widetilde{H}[\poiss^*_\lambda]$ graciously converge to
$\widetilde{h}[\norm]$ because now we respectively have
$\sigma=\sqrt{\lambda}$ and $\Delta x_k=\frac{1}{\sqrt{\lambda}}$

In a similar way for the discrete uniform distributions $\unif_n(a)$
introduced in the Section~\ref{intro} we now have
\begin{equation*}
    x_k=\frac{ka}{n}\quad\qquad\Delta x_k=\frac{a}{n}\quad\qquad
    p_k=\frac{1}{n}\quad\qquad k=1,\ldots,n
\end{equation*}
so that, since $\widetilde{\varrho}[\unif_n]$ is the \emph{IQrR}
$\varrho_\frac{1}{4}$ again, from\refeq{discrunifentropy}
and\refeq{newentropy} we get
\begin{equation*}
    \widetilde{H}[\unif_n]=\ln n
    +\ln\frac{a}{n}-\ln\widetilde{\varrho}[\unif_n]=\ln a-\ln\widetilde{\varrho}[\unif_n]
\end{equation*}
If we then remember from\refeq{unifiqr} that
$\widetilde{\varrho}[\unif_n]\ton\widetilde{\varrho}[\unif]=\frac{a}{2}$,
we finally get from\refeq{uniftilde}
\begin{equation*}
    \widetilde{H}[\unif_n]\ton\ln a-\ln\widetilde{\varrho}[\unif]=\ln
    a-\ln\frac{a}{2}=\ln 2=\widetilde{h}[\unif]
\end{equation*}
We are then allowed to conjecture that this is a generalized
behavior: within the frame of our definitions\refeq{newentropy} of
$\widetilde{H}$ and\refeq{newestdiffentropy} of $\widetilde{h}$,
whenever -- as in our previous examples -- a sequence of purely
discrete laws $\mathfrak{A}_n$ weakly converges to a continuous law
$\mathfrak{A}$, then also
$\widetilde{H}[\mathfrak{A}_n]\ton\widetilde{h}[\mathfrak{A}]$

\section{Conclusions}

We have proposed to modify both the usual definition\refeq{entropy}
of the entropy $H$, and\refeq{diffentropy} of the differential
entropy $h$ respectively into\refeq{newentropy}
and\refeq{newestdiffentropy}, namely within our most general
notation
\begin{eqnarray}
  \widetilde{H} &=&-\sum_{k\geq1}\frac{\Delta F_k}{\Delta x_k}\ln\left(\widetilde{\varrho}\,\frac{\Delta F_k}{\Delta x_k}\right)\Delta x_k=H-\ln\widetilde{\varrho}+\sum_{k\geq1}p_k\ln\Delta
    x_k\label{newentropy2}  \\
  \widetilde{h} &=&-\int_\RE f(x)\ln\left[\widetilde{\varrho}
    \,f(x)\right]\,dx=h-\ln\widetilde{\varrho}\label{newestdiffentropy2}
\end{eqnarray}
where in general $\widetilde{\varrho}$ coincides with the
\emph{IQrR} $\varrho\left(\frac{1}{4}\right)$ of the considered
distribution, except when the \emph{IQrR} vanishes (as can happen
for discrete laws): in this last event $\widetilde{\varrho}$ is
taken as the smallest non zero \emph{IQnR} of the distribution.
There are also several other possible re-definitions which
essentially differ among them by the choice of the parameter
$\kappa$ in\refeq{newdiffentropy2}, and by the set of their possible
values. All these definitions, moreover, bypass the anomalies listed
in the Section~\ref{intro} and appear to go smoothly one into the
other for suitable discrete-continuous limit. As a matter of fact
the introduction of the dimensional parameter $\kappa$ effectively
renormalizes the divergences that we would otherwise encounter in
the limiting processes leading from discrete to continuous laws.
Remark finally that the discrete form\refeq{newentropy2} can also be
easily customized to fit with the entropy estimation from empirical
data

We end the paper by pointing out that, despite extensive
similarities, the new quantities such as $\widetilde{H}$ and
$\widetilde{h}$ no longer have all the same properties of $H$ and
$h$. For instance at the present stage we could neither prove, nor
disprove (by means of some counterexample) that $\widetilde{H}\geq0$
as for $H$. On the other hand the examples seem also to allow no
room for $\widetilde{h}$-extremal distributions, as the normal laws
were for the $h$ differential entropy. Remark however that these
conclusions would be different by adopting the alternative
definitions that are presented in the Section~\ref{alternative}.
While all these topics seem to be interesting fields of inquiry, it
would also be important to extensively review what is preserved of
all the well known properties of the usual definitions, and how to
adapt further ideas such as relative entropy, mutual information and
whatever else is today used in the information
processing~\cite{cover,betten}. This remarks emphasize the
possibilities open by our seemingly naive changes: as a bid to
connect two previous standpoints, in fact, our proposed definitions
blend the properties of the older quantities, and in so doing can
also break new ground. An extensive analysis of all the possible
consequences both of the proposed definitions, and of their
articulations will be the subject of a forthcoming paper, while on
this topic we will at present limit ourselves just to point out that
many relevant features of the entropy essentially derive from the
properties of the logarithms which in any case play a central role
also in the new definitions. Finally it would be stimulating to
explore how -- if at all -- it is possible to make the new
definitions compatible with other celebrated extensions of the
classical entropies such as, for instance, that proposed by
Tsallis~\cite{tsallis}, or the more recent \emph{cumulative residual
entropy}~\cite{cre}

\begin{appendix}

\section{Examples}\label{examples}
\begin{figure}
 \begin{center}
\includegraphics*[width=8cm]{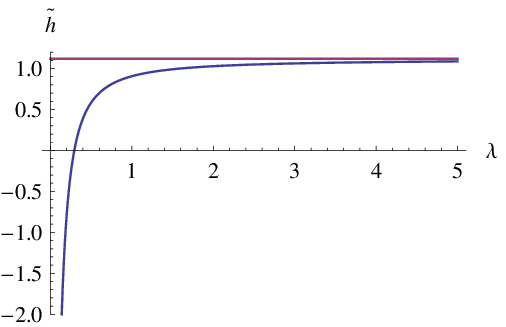}
\caption{Entropy $\widetilde{h}[\gam_\lambda]$ for gamma
laws}\label{gammatildefig}
 \end{center}
\end{figure}

We begin by comparing the values of the two differential entropies
$h$ and $\widetilde{h}$, respectively defined as\refeq{diffentropy}
and\refeq{newestdiffentropy}, for the most common families of laws
by neglecting the centrality parameters which are irrelevant for our
purposes because both the entropies are independent from them.

For the \textbf{Gaussian} laws $\norm(a)$ with
\begin{equation}\label{gauss}
    f(x)=\frac{e^{-\frac{x^2}{2a^2}}}{a\sqrt{2\pi}}\qquad\quad\sigma=a\qquad\quad
    h[\norm]=\ln\left(a\sqrt{2\pi e}\right)
\end{equation}
we know that
\begin{align}
  &  F(x)=\Phi\left(\frac{x}{a}\right)\quad\qquad\;\,\Phi(y)=\int_{-\infty}^y\frac{e^{-\frac{z^2}{2}}}{\sqrt{2\pi}}\,dz\label{gauss2}\\
  &  Q(p)=a\Phi^{-1}(p)\qquad\quad
  \widetilde{\varrho}\,[\norm]=a\left[\Phi^{-1}\left(\frac{3}{4}\right)-\Phi^{-1}\left(\frac{1}{4}\right)\right]\label{gauss3}
\end{align}
and hence we immediately have from\refeq{newestdiffentropy}
\begin{equation}\label{gausstilde}
    \widetilde{h}[\norm]=h[\norm]-\ln\widetilde{\varrho}\,[\norm]=\ln\frac{\sqrt{2\pi
    e}}{\Phi^{-1}\left(\frac{3}{4}\right)-\Phi^{-1}\left(\frac{1}{4}\right)}\approx1.11959
\end{equation}
For the laws $\unif(a)$ \textbf{uniform} on $[0,a]$ with (here
$\vartheta(x)$ is the Heaviside function)
\begin{equation}\label{unif}
    f(x)=\frac{\vartheta(x)-\vartheta(x-a)}{a}\qquad\quad\sigma=\frac{a}{\sqrt{12}}\qquad\quad
    h[\unif]=\ln a
\end{equation}
we instead have
\begin{equation}\label{unifiqr}
  F(x) = \frac{x}{a}\qquad  0\leq x\leq a \qquad\qquad
  Q(p) = ap\qquad 0<p<1\qquad\quad\widetilde{\varrho}=\frac{a}{2}
\end{equation}
so that
\begin{equation}\label{uniftilde}
    \widetilde{h}[\unif]=\ln a-\ln\frac{a}{2} =\ln 2\approx0.693147
\end{equation}
For the \textbf{gamma} laws $\gam_\lambda(a)$, $\lambda>0$ with
\begin{equation}\label{erl}
    f(x)=\vartheta(x)\frac{x^{\lambda-1}e^{-\frac{x}{a}}}{a^\lambda\Gamma(\lambda)}\qquad\quad\sigma=a\sqrt{\lambda}\qquad\quad
    h[\gam_\lambda]=(1-\lambda)\psi(\lambda)+\ln\left[ae^\lambda\Gamma(\lambda)\right]
\end{equation}
where
\begin{equation*}
    \psi(z)=\frac{\Gamma\,'(z)}{\Gamma(z)}
\end{equation*}
we have
\begin{align*}
  &  F(x)=\Gamma_\lambda\left(\frac{x}{a}\right)\qquad\quad\Gamma_\lambda(y)=
    1-\frac{\Gamma(\lambda,y)}{\Gamma(\lambda,0)}\qquad\quad\Gamma(\lambda,y)=\int_y^\infty
    t^{\lambda-1}e^{-t}dt \\
  & \qquad\qquad Q(p)=a\Gamma_\lambda^{-1}(p)\qquad\quad\widetilde{\varrho}=a\left[\Gamma_\lambda^{-1}\left(\frac{3}{4}\right)-\Gamma_\lambda^{-1}\left(\frac{1}{4}\right)\right]
\end{align*}
and hence
\begin{figure}
 \begin{center}
\includegraphics*[width=8cm]{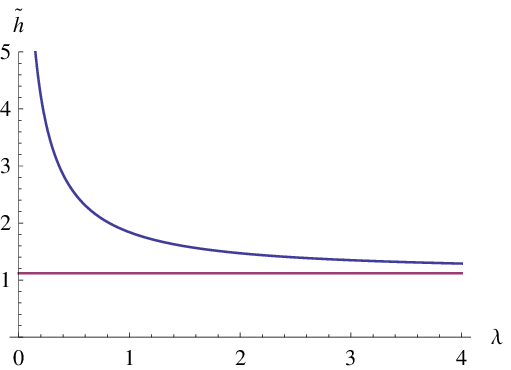}
\caption{Entropy $\widetilde{h}[\stud_\lambda]$ for Student
laws}\label{studenttildefig}
 \end{center}
\end{figure}
\begin{equation}\label{gammatilde}
    \widetilde{h}[\gam_\lambda]=(1-\lambda)\psi(\lambda)+\ln\frac{e^\lambda\Gamma(\lambda)}{\Gamma_\lambda^{-1}\left(\frac{3}{4}\right)-\Gamma_\lambda^{-1}\left(\frac{1}{4}\right)}
\end{equation}
which as a function of $\lambda$ is displayed in the
Figure~\ref{gammatildefig}: this shows that
$\widetilde{h}[\gam_\lambda]$ takes also negative values, that
$\widetilde{h}[\gam_\lambda]\to-\infty$ for $\lambda\to0$, and that
$\widetilde{h}[\gam_\lambda]\uparrow\widetilde{h}[\norm]$ for
$\lambda\to+\infty$. In particular for the \textbf{exponential} laws
$\esp(a)=\gam_1(a)$ with
\begin{equation}\label{esp}
    f(x)=\vartheta(x) \frac{e^{-\frac{x}{a}}}{a}\qquad\quad\sigma=a\qquad\quad
    h[\esp]=1+\ln a=\ln(e a)
\end{equation}
we have
\begin{equation*}\label{espgamma}
     F(x)=\left(1-e^{-\frac{x}{a}}\right)\vartheta(x)\qquad\quad Q(p)=-a\ln(1-p)\qquad 0<p<1
     \qquad\quad \widetilde{\varrho}=a\ln 3
\end{equation*}
so that
\begin{equation}\label{esptilde}
    \widetilde{h}[\esp]=1-\ln(\ln 3)\approx0.905952
\end{equation}
For the \textbf{Laplace} laws $\lapl(a)$ with
\begin{equation}\label{lapl}
    f(x)=\frac{ e^{-\frac{|x|}{a}}}{2a}\qquad\quad\sigma=a\sqrt{2}\qquad\quad
    h[\lapl]=\ln(2ae)
\end{equation}
we know that
\begin{equation*}
    \widetilde{\varrho}=2a \ln 2
\end{equation*}
and hence
\begin{figure}
 \begin{center}
\includegraphics*[width=8cm]{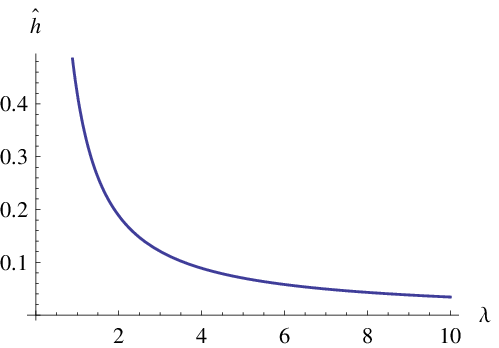}
\caption{Entropy $\widehat{h}[\gam_\lambda]$ for the gamma
laws}\label{gammahatfig}
 \end{center}
\end{figure}
\begin{equation}\label{lapltilde}
    \widetilde{h}[\lapl]=1-\ln\left(\ln 2\right)\approx1.36651\geq\widetilde{h}[\norm]
\end{equation}
Finally for the family of \textbf{Student} laws $\stud_\lambda(a)$,
$\lambda>0$ with
\begin{equation}\label{stud}
    f(x)=\frac{1}{aB\left(\frac{1}{2},\frac{\lambda}{2}\right)}\left(\frac{a^2}{a^2+x^2}\right)^\frac{\lambda+1}{2}
    \qquad\sigma=\frac{a}{\sqrt{\lambda-2}}\qquad\quad
    B(\alpha,\beta)=\frac{\Gamma(\alpha)\Gamma(\beta)}{\Gamma(\alpha+\beta)}
\end{equation}
the variance exists only for $\lambda>2$, but the differential
entropy is well defined for every $\lambda>0$
\begin{equation}\label{studentr}
    h[\stud_\lambda]=\frac{\lambda+1}{2}\left[\psi\left(\frac{\lambda+1}{2}\right)-\psi\left(\frac{\lambda}{2}\right)\right]+\ln\left[aB\left(\frac{1}{2},\frac{\lambda}{2}\right)\right]
\end{equation}
For short the explicit form of the \emph{IQrR} $\widetilde{\varrho}$
will not be explicitly given here. Since however the differential
entropy $h[\stud_\lambda]$, and the \emph{IQRr}
$\widetilde{\varrho}\,[\stud_\lambda]$ are both proportional to the
scaling parameter $a$, the entropy $\widetilde{h}[\stud_\lambda]$
will be independent from $a$, and as a function of $\lambda$ is
displayed in the Figure~\ref{studenttildefig} which shows that
$\widetilde{h}[\stud_\lambda]$ takes always positive values larger
than $\widetilde{h}[\norm]\approx1.11959$, that
$\widetilde{h}[\gam_\lambda]\to+\infty$ for $\lambda\to0$, and that
$\widetilde{h}[\gam_\lambda]\downarrow\widetilde{h}[\norm]$ for
$\lambda\to+\infty$. In particular for the \textbf{Cauchy} laws
$\cauchy(a)=\stud_1(a)$, without variance, with
\begin{align}
  &\qquad\qquad\quad  f(x)=\frac{1}{a\pi}\,\frac{a^2}{a^2+x^2}\qquad\quad
    h[\cauchy]=\ln\left(4\pi a\right)\label{cauchy} \\
  &  F(x)=\frac{1}{2}+\frac{1}{\pi}\arctan\frac{x}{a} \quad\qquad Q(p)=a\tan\frac{(2p-1)\pi}{2} \quad\qquad\widetilde{\varrho}=2a\label{cauchyiqr}
\end{align}
we have that
\begin{equation}\label{studenttilde}
     \widetilde{h}[\cauchy]=\ln(4\pi
     a)-\ln(2a)=\ln(2\pi)\approx1.83788\geq \widetilde{h}[\norm]
\end{equation}
A particular consequence of these example is that -- as already
remarked in the Section~\ref{interquant} -- the entropy
$\widetilde{h}$ has neither maximum, nor minimum value, and by
suitably choosing the continuous law it can take every real value
both positive and negative
\begin{figure}
 \begin{center}
\includegraphics*[width=8cm]{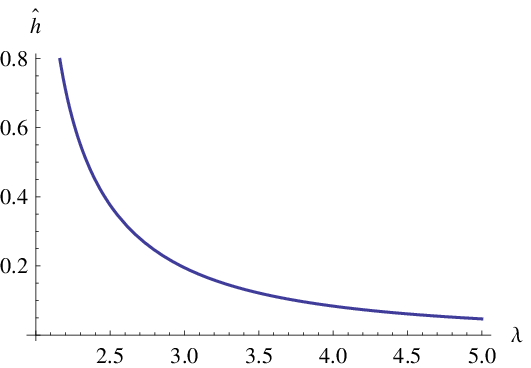}
\caption{Entropy $\widehat{h}[\stud_\lambda]$ for Student laws with
$\lambda>2$}\label{studentfig}
 \end{center}
\end{figure}

Similar calculations can then be carried on also for the alternative
entropy definition\refeq{newdiffentropy} of $\widehat{h}$ for laws
endowed with a finite second momentum: we immediately get for the
Gaussian laws
\begin{equation}\label{gausshat}
    \widehat{h}[\norm]=\ln\left(a\sqrt{2\pi e}\right)-\ln\left(a\sqrt{2\pi
    e}\right)=0
\end{equation}
and for the uniform laws
\begin{equation}\label{unifhat}
    \widehat{h}[\unif]=\ln\left(\frac{a}{\sqrt{12}}\sqrt{2\pi
    e}\right)-\ln a=\ln\sqrt{\frac{\pi
    e}{6}}\approx0.1765
\end{equation}
For the gamma laws we have now
\begin{equation}\label{gammahat}
    \widehat{h}[\gam_\lambda]=\ln\left(a\sqrt{2\pi e\lambda}\right)-(1-\lambda)\psi(\lambda)-\ln\left[ae^\lambda\Gamma(\lambda)\right]=\ln\frac{\sqrt{2\pi
    e\lambda}}{e^\lambda\Gamma(\lambda)}-(1-\lambda)\psi(\lambda)
\end{equation}
which always take positive values, as shown in the
Figure~\ref{gammahatfig}, with $\widehat{h}[\gam_\lambda]\to0$ for
$\lambda\to+\infty$, and in particular for the exponential law we
have
\begin{equation}\label{esphat}
    \widehat{h}[\esp]=\ln \left(a\sqrt{2\pi e}\right)-\ln (ea)=\ln\sqrt{\frac{2\pi}{e}}\approx0.4189
\end{equation}
For the Laplace laws we have
\begin{equation}\label{laplhat}
    \widehat{h}[\lapl]=\ln\left(a\sqrt{4\pi
    e}\right)-\ln(2ae)=\ln\sqrt{\frac{\pi}{e}}\approx0.0724
\end{equation}
and finally for the Student laws with $\lambda>2$
\begin{equation}\label{studentrhat}
    \widehat{h}[\stud_\lambda]=-\frac{\lambda+1}{2}\left[\psi\left(\frac{\lambda+1}{2}\right)-\psi\left(\frac{\lambda}{2}\right)\right]
    -\ln\left[\sqrt{\frac{\lambda-2}{2\pi e}}\,B\left(\frac{1}{2},\frac{\lambda}{2}\right)\right]
\end{equation}
displayed in the Figure~\ref{studentfig}. It is easy to check that
in all these examples the entropy $\widehat{h}$ takes only non
negative values
\begin{figure}
 \begin{center}
\includegraphics*[width=8cm]{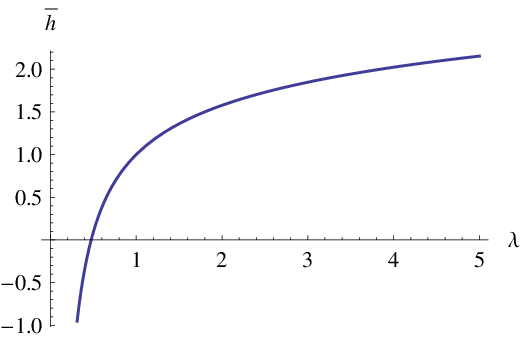}
\caption{Entropy $\overline{h}[\gam_\lambda]$ for gamma
laws}\label{gammabar}
 \end{center}
\end{figure}

\begin{figure}
 \begin{center}
\includegraphics*[width=8cm]{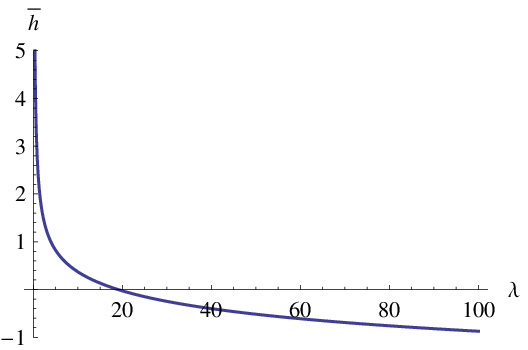}
\caption{Entropy $\overline{h}[\stud_\lambda]$ for Student
laws}\label{studentbar}
 \end{center}
\end{figure}

Finally for the third definition\refeq{typentropy} of $\overline{h}$
we have the following values: for the Gaussian type we have
\begin{equation}\label{typegauss}
    \overline{h}[\norm]=\ln\sqrt{2\pi e}\approx1.41894
\end{equation}
and for the uniform type
\begin{equation}\label{typeunif}
    \overline{h}[\unif]=0
\end{equation}
For the gamma types
\begin{equation}\label{typegamma}
    \overline{h}[\gam_\lambda]=(1-\lambda)\psi(\lambda)+\ln\left[e^\lambda\Gamma(\lambda)\right]
\end{equation}
the values, as displayed in the Figure~\ref{gammabar}, go from
$-\infty$ for $\lambda\to0^+$, to $+\infty$ for $\lambda\to+\infty$.
In particular for the exponential type we have
\begin{equation}\label{typesp}
    \overline{h}[\esp]=1
\end{equation}
For the Student types we finally have
\begin{equation}\label{typestud}
    \overline{h}[\stud_\lambda]=\frac{\lambda+1}{2}\left[\psi\left(\frac{\lambda+1}{2}\right)-\psi\left(\frac{\lambda}{2}\right)\right]+\ln\left[B\left(\frac{1}{2},\frac{\lambda}{2}\right)\right]
\end{equation}
with values shown in the Figure~\ref{studentbar} and going again
from $+\infty$ at $\lambda\to0^+$, to $-\infty$ for
$\lambda\to+\infty$. In particular for the Cauchy type we get
\begin{equation}\label{typecauchy}
    \overline{h}[\cauchy]=\ln(4\pi)\approx2.53102
\end{equation}
As remarked in the Section~\ref{alternative}, these examples show
that the entropy $\overline{h}$ takes again all the (positive and
negative) real values and hence that there is no maximum entropy
distribution as in the case of the $\widetilde{h}$ entropy

\section{Mixtures}\label{mixtures}

\begin{figure}
 \begin{center}
\includegraphics*[width=8cm]{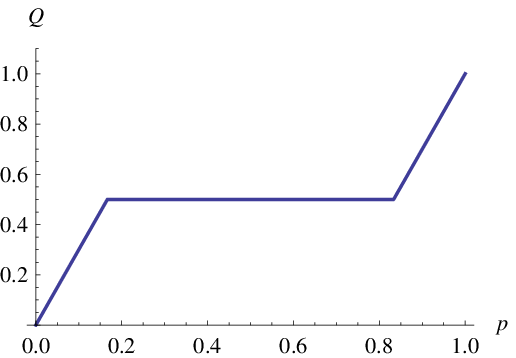}
\caption{The quantile function $Q(p)$ for a mixture of degenerate
and uniform laws}\label{mixturefig}
 \end{center}
\end{figure}

The definition of $\widetilde{\varrho}$ proposed in the
Section~\ref{discrlaws} certainly produces non zero values for both
purely discrete and purely continuous laws. Some care must be
exercised however for discrete-continuous mixtures. Let us take for
instance the mixture
\begin{equation*}
    \frac{2}{3}\,\bm\delta_\frac{1}{2}+\frac{1}{2}\,\unif(1)
\end{equation*}
of a law degenerate in $x=\frac{1}{2}$, and a law uniform in
$[0,1]$. Its \cdf\ would then be
\begin{equation*}
    F(x)=\left\{
           \begin{array}{ll}
             0 & \quad x<0 \\
             \frac{2}{3}\,\vartheta\left(x-\frac{1}{2}\right)+\frac{x}{3} & \quad 0\leq x\leq 1 \\
             1 & \quad 1< x
           \end{array}
         \right.
\end{equation*}
where $\vartheta$ is the Heaviside function, and its quantile
function is
\begin{equation*}
    Q(p)=\left\{
           \begin{array}{ll}
             3p & \quad 0\leq p\leq\frac{1}{6} \\
             \frac{1}{2} & \quad\frac{1}{6}\leq p\leq\frac{5}{6} \\
             3p-2 & \quad\frac{5}{6}\leq p\leq1
           \end{array}
         \right.
\end{equation*}
as displayed in the Figure~\ref{mixturefig}. It is apparent then
that the \emph{IQrR} is zero because
\begin{equation*}
    \varrho\left(\frac{1}{4}\right)=Q\left(\frac{3}{4}\right)-Q\left(\frac{1}{4}\right)=\frac{1}{2}-\frac{1}{2}=0
\end{equation*}
while the \emph{IQnR} $\varrho(p)$ is a continuous function so that
(with the notations of the Section~\ref{discrlaws}) also
$\widetilde{\varrho}=0$ because $\inf\mathcal{P}_0=0$. As a
consequence, in the case of discrete-continuous mixtures with a
\cdf\ such as
\begin{equation*}
    F(x)=qF_d(x)+(1-q)F_c(x)\qquad\quad 0<q<1
\end{equation*}
we can not simply extend the definitions of the
Section~\ref{discrlaws}. We could however consider separately both
the $\widetilde{\varrho}_d$ of the discrete distribution (as defined
in the Section~\ref{discrlaws}), and the
$\widetilde{\varrho}_c=\varrho\left(\frac{1}{4}\right)$ of the
continuous distribution, and to take as dimensional constant
$\kappa$ their convex combination
\begin{equation*}
    q\,\widetilde{\varrho}_d+(1-q)\widetilde{\varrho}_c
\end{equation*}
which never vanishes because at least its continuous part is always
non zero

\end{appendix}

\subsection*{Acknowledgements} The author would like to thank A.\
Andrisani, S.\ De Martino, S.\ De Siena, C.J.\ Ellison and S.\
Stramaglia for invaluable comments and suggestions


\begin{thebibliography}{99}

\bibitem{loeve} \emph{M.\ Lo\`eve}, \textsc{Probability Theory I--II} (Springer,
Berlin, 1977-8).

\bibitem{jacquet}
\emph{P.\ Jacquet, W.\ Szpankowski}, IEEE Trans. Inform. Th. 45
(1999) 1072

\bibitem{cichon} \emph{J.\ Cicho\'n, Z.\ Go{\l}\c{e}biewski}, DMTCS Proc.
AQ (2012) 179, 23rd Intern. Meeting on \emph{Probabilistic,
Combinatorial, and Asymptotic Methods for the Analysis of
Algorithms} (AofA'12)

\bibitem{cover} \emph{T.M.\ Cover, J.M.\ Thomas}, \textsc{Elements of Information
Theory}, (Wiley, Hoboken, New Jersey, 2006)

\bibitem{betten} \emph{L.M.A.\ Bettencourt, V. Gintautas, M.I.\ Ham},
Phys.\ Rev.\ Lett. \textbf{100}, 238701 (2008)

\bibitem{tsallis} \emph{C.\ Tsallis}, J.\ Stat.\ Phys.\ \textbf{52} (1988) 479

\bibitem{cre} \emph{N.\ Drissi, T.\ Chonavel, J.M.\ Boucher}, Res.\ Lett.\ Signal
Proc.\ (2008) ID 790607


\end{thebibliography}
\end{document}